\documentclass{article}
\usepackage{graphicx} 
\usepackage{phase1}
\usepackage[square,sort,comma,numbers]{natbib}
\usepackage{aas_macros}
\usepackage[colorlinks,backref=page]{hyperref}
\usepackage[table,xcdraw,svgnames]{xcolor}
\usepackage{graphicx} 
\hypersetup{
    colorlinks=true,
    linkcolor=magenta,
    citecolor = blue,
    filecolor=magenta,      
    urlcolor=blue,
    pdftitle={Overleaf Example},
    pdfpagemode=FullScreen,
    }

\renewcommand*{\backref}[1]{}
\renewcommand*{\backrefalt}[4]{%
    \ifcase #1 %
        (no citations)%
    \or
        (back to page #2)%
    \else
        (cited on pages #2)%
    \fi
}
\begin{document}


{\large \begin{center} \textbf{Building a Roadmap for Hubble science into the 2030s\\``Crucial UV spectroscopy of Oe stars in nearby galaxies''}\end{center}}


{\normalsize \begin{center} A. Wofford$^{1}$,  D. Pauli$^{2}$, A.A.C.\ Sander$^{3}$, O. Aranguré$^{1}$, M.S. Oey$^{5}$, O.G. Telford$^{6}$, P.A.\ Crowther$^{7}$, Jorick S. Vink$^{8}$, Tomer Shenar$^{9}$, M. Gull$^{10,11}$, L.P. Martins$^{12}$, S. Simón-Díaz$^{13,14}$, S. Zharikov$^{1}$, L.\,M. Oskinova$^{15}$, L.J. Smith$^{16}$

\end{center}}

{
\scriptsize \begin{center}
$^{1}$Universidad Nacional Autónoma de México. Instituto de Astronomía. A.P. 106, 22800. Ensenada, B.C., México\\
$^{2}$Institute of Astronomy, KU Leuven, Celestijnenlaan 200D, 3001 Leuven, Belgium\\
$^{3}$Zentrum f{\"u}r Astronomie der Universit{\"a}t Heidelberg, Astronomisches Rechen-Institut, M{\"o}nchhofstr. 12-14, 69120 Heidelberg, Germany\\
$^{5}$University of Michigan, Department of Astronomy, 1085 South University Avenue, Ann Arbor, MI 48109-1107, USA\\
$^{6}$Department of Physics and Astronomy, University of Utah, 275 South University Street, Salt Lake City, UT 84112, USA\\
$^{7}$Astrophysics Research Cluster, Mathematical \& Physical Sciences, University of Sheffield, Sheffield S3 7RH, UK\\
$^{8}$Armagh Observatory and Planetarium, College Hill, BT61 9DG Armagh, Northern Ireland\\
$^{9}$The School of Physics and Astronomy, Tel Aviv University, Tel Aviv 6997801, Israel\\
$^{10}$The Observatories of the Carnegie Institution for Science, 813 Santa Barbara Street, Pasadena, CA 91101, USA\\
$^{11}$Department of Astronomy, California Institute of Technology, Pasadena, CA 91125, USA\\
$^{12}$ N\'ucleo de Astrofı\'isica, Universidade Cidade de S\~ao Paulo, Rua Galv\~ao Bueno 868, SP 01506-000, Brazil\\
$^{13}$Instituto de Astrofísica de Canarias, C. Vía Láctea, s/n, 38205 La Laguna, Santa Cruz de Tenerife, Spain
\\
$^{14}$Universidad de La Laguna, Dpto. Astrof\'isica, Av. Astrofs\'ico Francisco S\'anchez, 38206 La Laguna, Santa Cruz de Tenerife, Spain\\
$^{15}$ University of Potsdam, Institute of Phsyscs and Astronomy, Germany\\
$^{16}$Space Telescope Science Institute, 3700 San Martin Drive, Baltimore, MD 21218, USA

\end{center}}

  \vspace*{-6pt}
    \begin{mdframed}[linecolor=black,linewidth=1pt]
    Hubble's unique COS G130M+G160M and STIS E140M UV spectral capabilities are essential for characterizing and understanding fundamental properties of main-sequence O-type emission-line (Oe) stars. These are fast rotators, and some are believed to be spun up in binaries. UV medium resolution observations of these stars are crucial for understanding massive binaries and their role in galaxy evolution. Oe stars are more prevalent at low metallicity, where they are highly under-studied, but UV spectra of these stars at all metallicities are needed. Observations of these stars in the 2030's with Hubble are particularly important in the era of ultra wide-field IFU optical and transient astronomy surveys. Ultimately, these observations will inform future UV observations with the Habitable Worlds Observatory. 
\noindent  
         \end{mdframed}
  \vspace*{-4pt}

\section{What are Oe stars, why are they important, and why is HST essential to study them?}
Understanding  massive ($\gtrsim8\,M_\odot$) O- and B-type stars
and their evolution 
is crucial for 
understanding feedback effects that drive galaxy evolution, including
ionizing radiation, mechanical energy, 
and the production of metals, cosmic rays, and dust.
They are also the progenitors of time-domain and multi-messenger phenomena including
supernovae, gamma-ray bursters, ultraluminous X-ray sources, and gravitational wave events.
Binary mass transfer is now known to affect the vast majority of massive stars \citep[e.g.,][]{Marchant23, Sana25}. Mass transfer removes mass from the donor and causes momentum transfer that spins up the companion, which can lead to OBe stars. Mass and momentum transfer greatly complicate stellar evolution of these stars and their populations, and it is strongly sensitive to metallicity \citep[e.g.,][]{Klencki2022A&A...662A..56K}.

OBe stars are a large and easily identified population of OB stars with Balmer emission from decretion disks caused by their extreme rotational velocities \citep[e.g.,][]{Rivinius2013A&ARv..21...69R}.  We expect that rapidly rotating stars become quasi chemically homogeneous, which increases the effective temperatures (T$_{\rm eff}$), lifetimes, and ionizing-photon production rates of OB stars; and to change their evolutionary tracks and outcomes \citep[e.g.,][]{Ekstrom2008PhDT........65E, Abdul-Masih2023A&A...669L..11A,Brott2011A&A...530A.115B}.

Most OBe stars are believed to be spun up as mass gainers in binaries \citep[e.g.,][]{Klement2019, Bodensteiner2020}, and therefore, OBe stars are a crucial population for understanding massive binary evolution.
Oe stars are Be stars on steroids. They are more prevalent at low metallicity, where they can constitute a significant fraction of the O stars (12-30\%) and also
reach earlier (hotter) types \citep[e.g.,][]{Golden-Marx2016ApJ...819...55G, Shenar2024A&A...690A.289S}, thus contributing significantly to Lyman continuum emission. 


Unfortunately, when line emission fills in their key optical hydrogen Balmer and He I absorption lines, it is extremely challenging to infer their spectral type, T$_{\rm eff}$, and ionizing spectrum using conventional optical analysis. 



\begin{figure}[h]
    \centering
    \includegraphics[width=\linewidth]{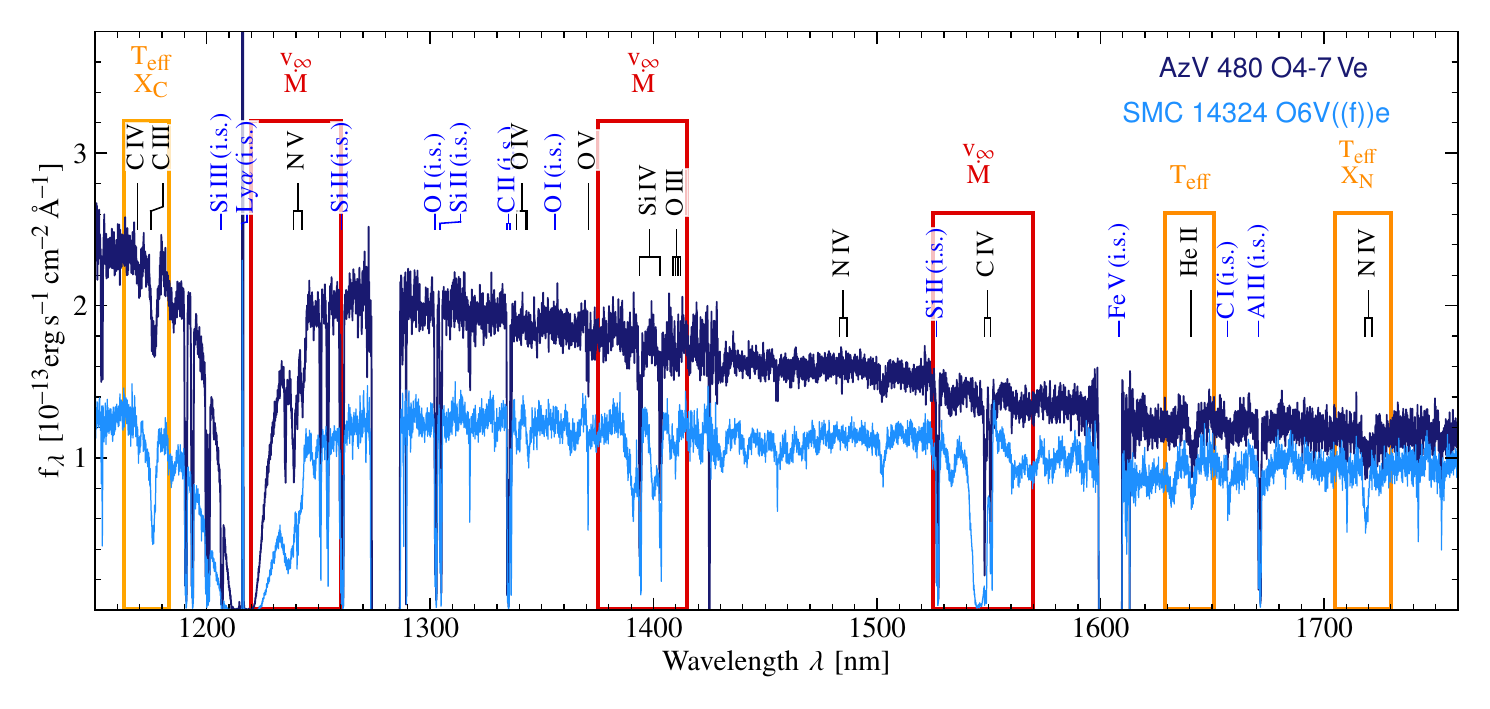}
    \caption{Example of two SMC Oe star spectra from the ULLYSES sample (blue lines). These spectra contain plenty of crucial information about the stellar (orange boxes) and wind parameters (red boxes). Fitting synthetic stellar atmosphere models to UV spectra like these ones is the only way to measure the temperatures and CNO abundances of Oe stars, since they are not contaminated by disc features. Furthermore, by fitting the N\,V, Si\,IV, and C\,IV lines provides crucial information on the mass-loss rates of mass gainers, a important step in understanding stellar wind physics. Credit: D. Pauli and ULLYSES team.}
    \label{fig:Oe_spectra}
\end{figure}
\begin{figure}[h]
    \centering
\includegraphics[width=0.9\linewidth]{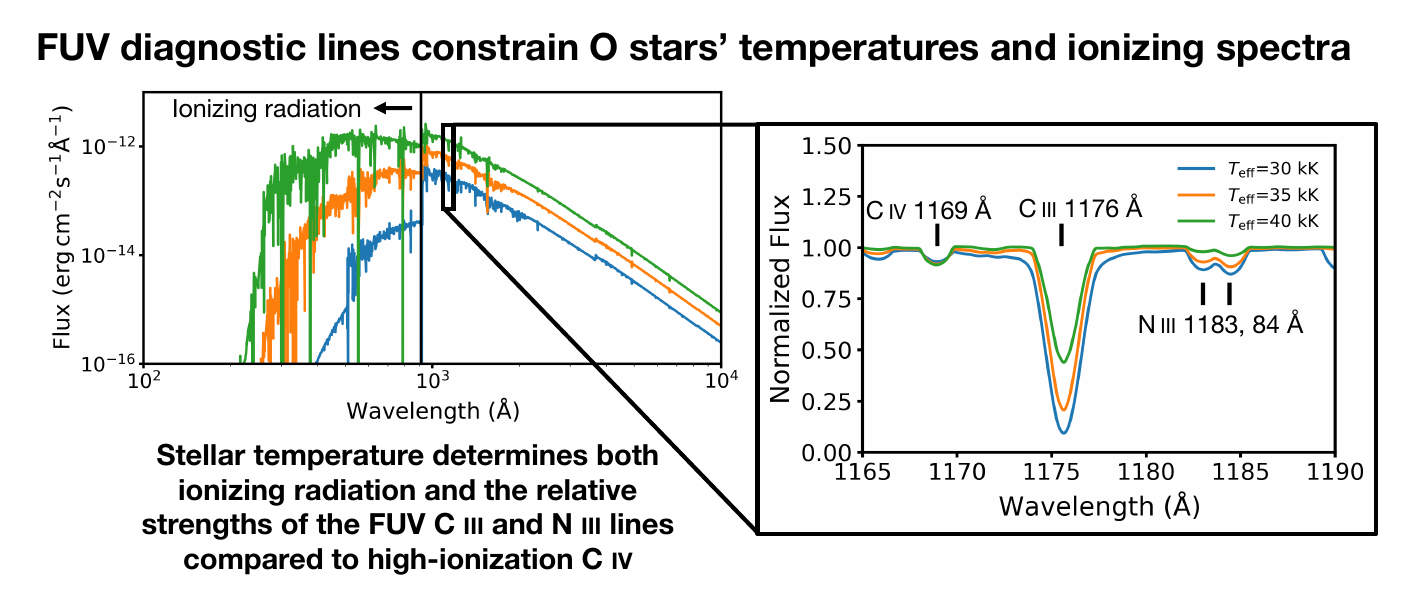}
    \vspace{-15pt}
    \caption{PoWR models of fast-rotating O stars at three different temperatures. {\bf Left:} hotter stars produce more and harder ionizing flux. {\bf Right:} zoom-in to temperature-sensitive FUV lines. FUV  observations are required to accurately predict the Oe stars' ionizing spectra. Credit: O.\ G.\ Telford.}
    \label{fig:fuv}
\end{figure}

FUV spectra offer the only route for constraining the fundamental properties of Oe stars (spectral type, stellar wind parameters, T$_{\rm eff}$, Q$_{\rm H}$). In Fig.~\ref{fig:Oe_spectra}, we show HST Oe-star observations and highlight important UV spectral features useful for deriving their main properties. In particular, for Oe stars, rest-frame FUV spectra offer an alternate way to constrain T$_{\rm eff}$ via the strengths of C III 1176 or N III 1183,84, relative to C IV 1169; and are essential to model ionizing spectra (see Fig.~\ref{fig:fuv}). This is why UV observations of Oe stars with HST are essential.

At low metallicity, only the following Oe stars are part of the recent HST-FUV Director’s Discretionary
program ULLYSES or its archival component   \citep{Roman-Duval2020RNAAS...4..205R, Bestenlehner2025arXiv250117931B}; and have available or planned optical IFU data. LMC: VFTS 102, BI 184 and BI 189 (all 3 in crowded areas); and Sk -70 60 (has a very faint nebula). SMC: AzV480 and AzV493 (both have very faint nebulae); and SMC 14324, also known as OGLE J004942.75-731717.7 (in a crowded area, shown in Fig.~\ref{fig:Oe_spectra}). Few FUV spectra of Oe stars are available, and most of them are of stars in complex regions that prevent us from using their associated nebulae to validate their ionizing fluxes. While the sample is small, it has become evident that the stellar and wind parameters of these stars are fundamentally different (e.g., see the C IV resonance line in Fig. 1). 

\begin{figure}[h]
    \centering
\includegraphics[width=1\linewidth]{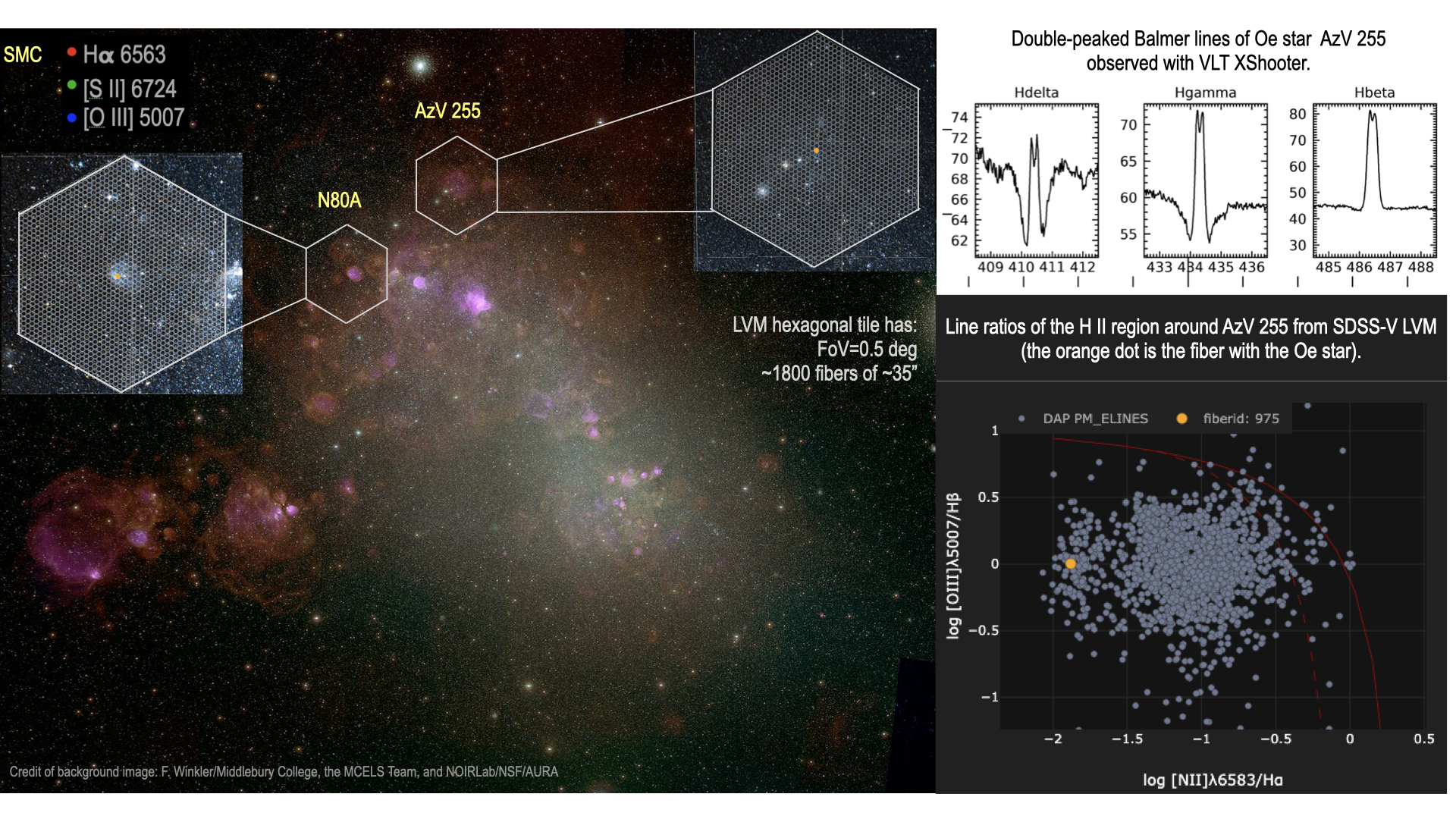}
    \vspace{-30pt}
    \caption{The ionizing fluxes of Oe stars can be validated using the luminosity of H II regions in the the Balmer lines. {\bf Left:} MCELS composite image with hexagonal LVM footprints  centered on Oe stars N80A and AzV 255. The orange dot is the fiber containing each target. {\bf Right-top:} For  AzV 255, we show the double-peaked Balmer lines observed as part of the XShootU program \cite{Vink2023A&A...675A.154V, Sana2024A&A...688A.104S}. Similar optical observations from Magellan are available for N80A. {\bf Right-bottom:} For the same star, we show a BPT \citep{Baldwin1981PASP...93....5B} diagnostic diagram where each point corresponds to one of the LVM fibers in the tile containing AzV 255. The orange dot is AzV 255. The faint red line separates regions ionized by massive stars from AGN ionization. Credit: A. Wofford and MCELS, XShootU, and SDSS V LVM teams.}
    \label{fig:keyfig}
\end{figure}

In order to validate the ionizing fluxes of Oe stars inferred from their stellar FUV spectra, wide field IFU surveys of the ionized-gas of nearby galaxies, like the ``Local Volume Mapper'' (LVM) of the Sloan Digital Sky Survey-V are extremely useful. LVM is producing an unprecedented spatially-resolved study of the nebular gas across large portions of nearby galaxies, including the Small Magellanic Cloud (metallicity $Z \approx Z_\odot/5$). It  covers from 3,600 to 10,000 \AA~at $\rm{R}\approx4,000$, with a  resolution of $\approx10$ pc at the distance of the SMC  \citep{Drory2024AJ....168..198D}. The LVM data will become public in June 2027, during HST Cycle 34. LVM coverage is useful for validating the ionizing fluxes inferred from the stellar FUV spectra, but not sufficient to calibrate stellar atmosphere models without the UV. In Fig.~\ref{fig:keyfig}, we show the location of two isolated Oe stars in the SMC which currently do not have FUV spectra and are covered by LVM. They were easily identified via their characteristic doubled-peaked Balmer emission lines, shown for one of the targets in  Fig.~\ref{fig:keyfig}. In order to take full advantage of spectroscopic surveys of nebular gas like the LVM, which targets the Southern MW and Southern star-forming galaxies in the Local Group, it is essential to constrain the ionizing spectra of Oe stars in these galaxies, via HST observations.   

\section{Key science questions which require Hubble’s capabilities}

\subsection{What are fundamental stellar properties of Oe stars?}

Determining the effective temperature of Oe stars is essential for estimating the properties of the ionizing spectra of Oe stars. These spectra are essential for modeling the spatially-resolved properties of their surrounding H II regions (density and temperature structure, chemical abundances, and spectral properties, e.g., \cite{Simon-Diaz2008MNRAS.389.1009S, Simon-Diaz2011A&A...530A..57S,Telford_2023}). This information is also essential for understanding the dynamics of H II regions, which could be dominated by radiation pressure, and the role of stellar feedback on subsequent star formation \citep[e.g.][]{Krumholz2009ApJ...703.1352K}. Finally, 
the leakage of ionizing photons locally and from their host galaxies, 
is central to understanding the diffuse ionized gas (DIG) of the ISM, 
the circumgalactic medium, and cosmic reionization. Figure~\ref{fig:keyfig}, illustrates the importance of determining reliable ionizing fluxes for Oe stars in the context of ionized-gas interstellar studies.

Oe stars are {\it ubiquitous at low metallicity} \citep[e.g.,][]{Golden-Marx2016ApJ...819...55G}. It is crucial to understand the dependence of their ionizing spectra with metallicity by observing targets spanning a range of metallicities, such as the Milky Way, Magellanic Clouds and other nearby galaxies. 

\subsection{
Do Oe stars have companions and if so, what are their characteristics 
and orbital properties?}

In the last decade, OBe stars have been identified as potentially crucial stepping stones in the binary evolution path to compact binaries, and ultimately gravitational wave events \citep[e.g.,][]{Langer2020,Marchant23, Schuermann2025}. In this scenario, OBe stars have exotic companions such as stripped stars, neutron stars, and black holes, which may impact the ionizing budget of low-metallicity galaxies. Studying OBe stars in the UV provides significant constraints on the nature of companions, if present.

At low metallicity, 
binary interactions and other mechanisms that may generate OBe stars remain poorly empirically constrained. Additionally, there is still uncertainty in which binary parameter space (e.g., mass ratio, orbital separation, etc.) corresponds to the OBe progenitors. These are central to constraining the
the properties and evolution of massive binary populations \citep[e.g.][]{Pols1991, Schuermann2025}.



\section{Instrument capabilities and operational requirements}

Ideally, two medium resolution COS gratings are needed to cover temperature sensitive photospheric lines such as C III 1169 and C IV 1176 (G130M) and stellar wind lines such as C IV 1550, He II 1640, and N IV 1718 (G160M). In addition, COS G140L is useful for stars with higher reddening, needing larger exposure times. Conversely, STIS E140M and E230M are needed for stars which are too bright for COS or in more densely populated regions. 

\section{Towards the development of the HWO}

HWO would enable to obtain UV spectra of Oe stars in Local Group galaxies and beyond. In order to help plan these observations, it is crucial to know the properties of a representative sample of local Oe stars, via HST UV spectroscopy.

\section{Hubble Legacy Programs}

We foresee a community driven Legacy proposal focused on obtaining UV spectra of a sample of optically identified Oe stars or Oe star candidates in the Milky Way, the Magellanic Clouds, and other Local Group galaxies, such as IC 1613, M33, and possibly NGC 6822.

\bibliographystyle{aasjournal}
\bibliography{references.bib}{}
\end{document}